# Colossal Seebeck coefficient in Aurivillius Phase-Perovskite Oxide Composite

Ashutosh Kumar[1,*,#], D. Sivaprahasam[2], Ajay D. Thakur[1,*]

[1]Department of Physics, Indian Institute of Technology Patna, Bihta 801 106, India
[2]Center for Automotive Energy Materials, ARC International IITM Research Park, Chennai 600 113 India

**Abstract:** We propose an inexpensive, scalable approach for achieving extremely high values of the Seebeck coefficient ($\alpha$) by exploiting the natural superlattice structure in Aurivillius phase oxides. In particular, we report an $\alpha \approx 319$ mV/K at 300 K in a composite of Aurivillius phase compound $SrBi_4Ti_4O_{15}$ (as a matrix) and a perovskite phase material (e.g., $La_{0.7}Sr_{0.3}MnO_3$ or, $La_{0.7}Sr_{0.3}CoO_3$ as filler). Such a colossal value of $\alpha$ may be attributed to contributions from the enhanced density of states due to the effective low dimensional character of the $Bi_2O_2$ layer. The corresponding thermal conductivity ($\kappa$) and the electrical conductivity ($\sigma$) lies in the range 0.7 - 1.25 W/m-K and 10 - 100 µS/m, respectively at 300 K. Attributed to the high $\alpha$ values, such oxide composites can be used as suitable materials for thermopile sensors and highly sensitive bolometric applications. We anticipate that the demonstration of colossal $\alpha$ in oxide composites using a simple synthesis strategy also sets the stage for future material innovations for high temperature thermoelectric applications.

*Corresponding author(s): science.ashutosh@gmail.com; ajay.thakur@iitp.ac.in

#Present address: Lukasiewicz Research Network-Cracow Institute of Technology Krakow, Poland

# INTRODUCTION

Thermoelectric (TE) materials have found practical applications in energy harvesting, solid-state cooling, and power supplies in deep space probes[1][2][3]. The efficient working of such applications requires a high value of Seebeck coefficient or thermopower ($\alpha$), a high electrical conductivity ($\sigma$), a low thermal conductivity ($\kappa$). However, the interdependence of these parameters leads to severe constraints in finding TE materials suitable for applications [4]. Several strategies, including band-gap engineering [5], modulation doping [6], energy filtering [7], and phonon glass electron crystal (PGEC)[8][9] have been suggested in recent past to circumvent these limitations.

In this manuscript, we focus on achieving the colossal values of $\alpha$. This is expected to set the stage for future material innovations that may lead to the optimization of the other crucial parameters, viz., $\sigma$, and $\kappa$ leading to high-quality thermoelectric materials. The choice of oxide systems for such a purpose provides the added advantage of stability at high temperatures for practical applications. Due to their high $\alpha$ values, these materials can be useful for bolometric applications and as thermopile and infrared sensors [10][11]. Several approaches for enhancing $\alpha$ have been proposed in the past. The coupling of the quantum of lattice vibrations (phonons) with charge carriers has been propounded as one of the key mechanisms for achieving high $\alpha$ at low temperatures [12][13][14]. The underlying phenomenon of the transfer of momentum from the non-equilibrium phonons to charge carriers is known as phonon drag (PD). It leads to an additional contribution $\alpha_{pd}$ to the typical diffusion contribution to thermopower, $\alpha_d$ [15]. Carrier energy filtering can be implemented using the following strategies: (a) tuning the scattering of electrons based on their energies by the introduction of quantum dots in the host matrix [16][17][18] and (b) scattering of low energy electrons by introducing an energy barrier at the interfaces [19][20]. Hicks and Dresselhaus proposed quantum confinement as a route to enhance $\alpha$ using low dimensional channel by exploiting the enhancement in the density of states, with minimal reduction in $\sigma$ [21][22] and was shown experimentally in several superlattice systems [23][24]. However, making superlattice structures require expensive fabrication facilities. Besides, the maximum reported value of $\alpha$ in superlattices are of the order of several hundreds of µV/K [25]. Koumoto et al. proposed a possibility to improve the $\alpha^2\sigma$ in quantum nanostructured bulk $SrTiO_3$ ceramic with two-dimensional electron gas (2DEG) grain boundaries (GBs) in terms of brick and mortar in three dimensional systems. The 2DEG formed at the GBs gives rise to the energy filtering effect that improves $α^2σ$ [26]. Large values of α were reported in several systems [27][28][29][30].

In literature, several exciting TE properties were reported in oxide systems with a layered structure, e.g., $NaCo_2O_4$ [31], $Ca_3Co_4O_9$ [32] Recently, H. Kohri et al. showed a significantly high value of α in the $Bi_2VO_{5.5}$ system: an Aurivillius phase (AP) material [33][34]. This class of materials are potential oxygen ion conductors [35]. The presence of a natural superlattice structure in the AP system is quite interesting for α and κ [36]. Nevertheless, these systems were ignored for TE applications due to their poor σ. Electrical conductivity in a TE system may be improved either by substitution of elements having different charge states as well as with the addition of conducting the second phase [37][38][39]. We propose an inexpensive, scalable approach for achieving extremely high values of α by exploiting the natural superlattice structures in an Aurivillius phase oxide material [40]. In the present study, we report a systematic study of thermoelectric properties in $SrBi_4Ti_4O_{15}$ (SBTO) AP-$La_{0.7}Sr_{0.3}MnO_3$/$La_{0.7}Sr_{0.3}CoO_3$ perovskite composite over a wide temperature range from 300 K-800 K. Further, colossal α obtained in the present study is used to demonstrate the thermopile sensor application.

## EXPERIMENTAL SECTION

$SrBi_4Ti_4O_{15}$ (SBTO), $La_{0.7}Sr_{0.3}MnO_3$ (LSMO) and $La_{0.7}Sr_{0.3}CoO_3$ (LSCO) were synthesized using a standard solid-state route (SSR), as mentioned in our previous reports [41][42]. The AP-perovskite composites (SBTO+x wt% LSMO/LSCO) were prepared by mixing the LSMO/LSCO powder with SBTO in a certain weight percentage (x). The structural characterization is performed using a powder x-ray diffraction (PXRD) technique, followed by Rietveld refinement. The composite samples were sintered using a conventional sintering process (@ 800°C for 6 hours with 3°/min cooling and heating rate) as well as using spark plasma sintering (SPS) at 800°C for 5 minutes with 70 °C/min heating and cooling rate. The σ and α are measured simultaneously using the standard four-probe method. The κ of the composite was measured using the following equation: κ = $Dc_p\rho$. The thermal diffusivity (D) was measured using laser-flash analysis, specific heat capacity ($c_p$) was calculated using Dulong-Petit law, and the sample density (ρ) was measured using the sample mass and its geometrical volume.

**RESULTS AND DISCUSSION**

Figure 1(a) depicts the PXRD patterns for SBTO and SBTO-LSMO/LSCO composites. The diffraction pattern due to SBTO confirms the pure phase formation with the characteristics 2θ peak at 30.38° and is in agreement with the ICDD File No. 043-0973. In the composite samples, the PXRD pattern due to SBTO and LSMO/LSCO is observed, and no impurity peak is detected within the sensitivity of the PXRD. The intensity corresponding to LSMO/LSCO is found to increase with the increase in the LSMO/LSCO concentration in the composite.

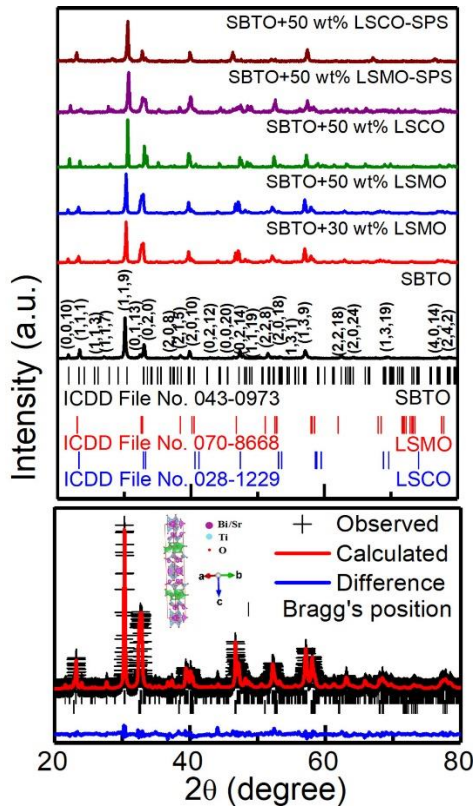

FIG. 1: (a) X-ray diffraction pattern for $SrBi_4Ti_4O_{15}$ + x wt% $La_{0.7}Sr_{0.3}MnO_3$/$La_{0.7}Sr_{0.3}CoO_3$ composite. Bragg peaks corresponding to SBTO (black), LSMO (red), LSCO (blue) are marked. (b) Rietveld refinement

pattern of SrBi$_4$Ti$_4$O$_{15}$ + 50 wt% La$_{0.7}$Sr$_{0.3}$MnO$_3$. The inset shows the structure of a natural superlattice Aurivillius phase (SrBi$_4$Ti$_4$O$_{15}$) using the atomic position from the Rietveld refinement.

Further, two-phase Rietveld refinement of the PXRD pattern of the composites is done using FullProf$^{TM}$ software, and refinement pattern for SBTO+50 wt%LSMO is shown in Fig. 1(b). In the refinement, LSMO/LSCO is taken as a rhombohedral structure (R-3c) and SBTO as a tetragonal structure (A21am). The goodness of fit ($\chi^2$) for SBTO+30 wt%LSMO, SBTO+50 wt%LSMO, and SBTO+50 wt%LSCO is 1.84, 1.92, and 1.79, respectively. It is worth noting that the phase fraction obtained from the two-phase Rietveld refinement is in agreement with the nominal composition of the composites. The obtained weight fraction, refinement parameters, along with the lattice parameters for SBTO, LSMO, and LSCO, are shown in Table I.

TABLE I: The weight fraction and $\chi^2$ obtained from the two-phase Rietveld refinement for the SBTO + x wt% LSMO/LSCO composite. The following lattice parameters are used for the refinement: LSMO (a=5.462 Å, c=13.147 Å), LSCO (a=5.444 Å, c=13.205 Å and SBTO is a=5.447 Å, c=41.165 Å).

| Sample composition | $\chi^2$ | weight fraction | | | |
| --- | --- | --- | --- | --- | --- |
| | | SBTO (%) | | LSMO (%) | |
| | | Nominal | Obtained | Nominal | Obtained |
| SBTO | 1.90 | 100 | 100 | 0 | 0 |
| SBTO+30 wt%LSMO | 1.84 | 76.92 | 80.15 | 23.07 | 20.85 |
| SBTO+50 wt%LSMO | 1.92 | 66.67 | 68.24 | 33.33 | 31.76 |
| SBTO+50 wt%LSCO | 1.79 | 66.67 | 69.17 | 33.33 | 30.83 |
| SBTO+50 wt%LSMO-SPS | 1.91 | 66.67 | 69.47 | 33.33 | 30.53 |
| SBTO+50 wt%LSCO-SPS | 1.95 | 66.67 | 69.35 | 33.33 | 30.65 |

The surface morphology of the SBTO+50 wt%LSMO, SBTO+50 wt%LSCO, and their SPS samples are shown in Fig. 2(a-d). In the non-SPS sample, several pores, along with grains of both the phases, are observed in the surface morphology. However, in the SPS sample, these pores and small grains are eliminated and show a highly compact structure. Fig. 2(e) shows the TEM image of the SBTO+50 wt%LSMO composite sample, which further confirms the presence of two different lattice spacing in the composite. The two different inter-planner distance, estimated using ImageJ software, corresponds to LSMO ≈ 0.274 nm(104) (Fig. 2(f)) and SBTO ≈ 1.041 nm (004) (Fig. 2(g)) phases. The inter-planner distance corresponding to SBTO is large in comparison with LSMO, as evident from large lattice parameters of SBTO compared to LSMO.

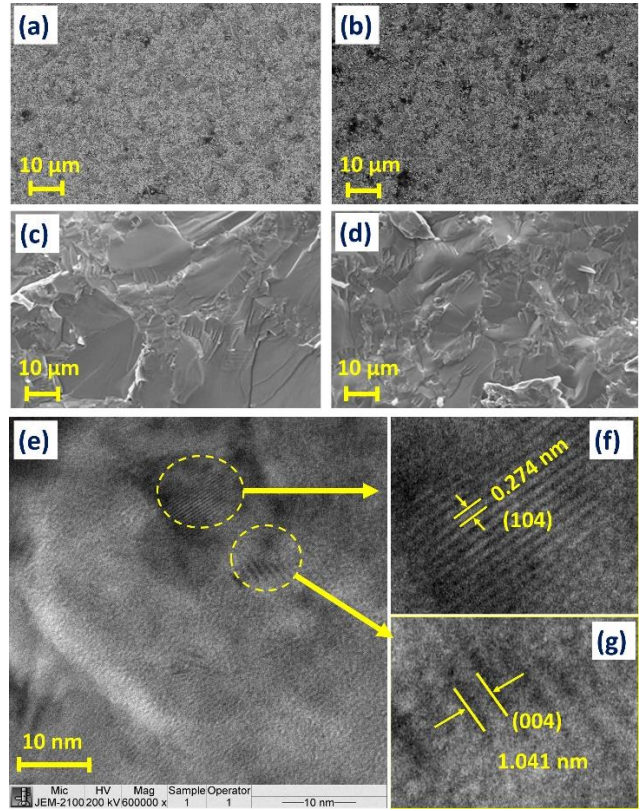

FIG. 2: FESEM images of (a) SBTO+50 wt% LSMO (b) SBTO+50 wt% LSCO (c) SBTO+50 wt% LSMO-SPS (d) SBTO+50 wt% LSCO-SPS. (e) TEM image for SBTO+50 wt% LSMO-SPS sample shows the existence of lattice planes corresponding to LSMO (f) and SBTO (g) phases in the composite.

The pure AP material, SBTO, is highly resistive, and it is difficult to measure its $\sigma$ and $\alpha$. The LSMO and LSCO are used as a conductive second phase to improve the overall electrical conductivity of the SBTO-LSMO/LSCO composite. The electrical conductivity of LSMO is ~500 S/cm [43] at 300 K. In contrast, the electrical conductivity for LSCO is 1000 S/cm [44] at the same temperature, which is quite higher than the electrical conductivity of SBTO. Hence the electrical conductivity of the composite with LSCO shows more electrical conductivity than that of LSMO. When LSMO/LSCO is mixed with SBTO, the resistivity of the composite decreases, and after a certain weight percentage of LSMO/LSCO, the sample becomes measurable. The $\sigma$ and $\alpha$ measurement of the composite samples from 300 K-800 K is shown in Fig. 3(a).

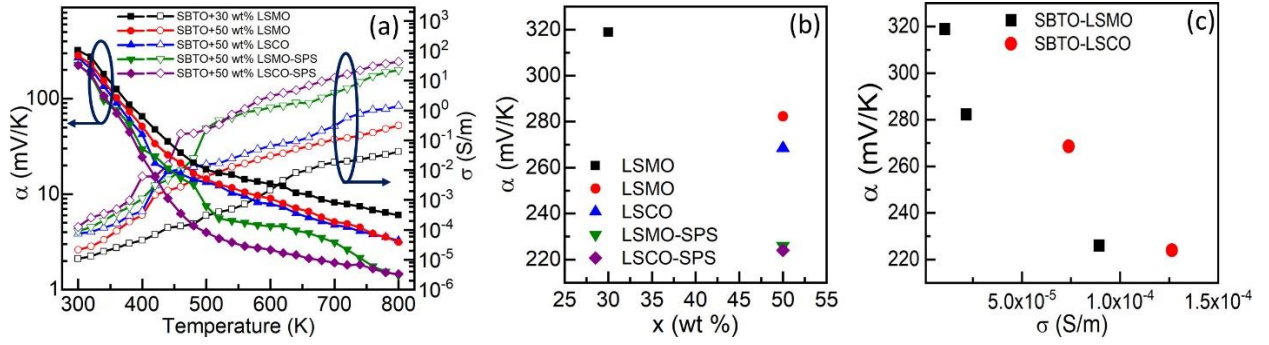

FIG. 3: (a) Temperature variation of Seebeck coefficient (α) marked with filled symbols and electrical conductivity (σ) marked with open symbols, (b) Seebeck coefficient (α) as a function of weight fraction (x) in the for SBTO+x wt% LSMO/LSCO composites, and (c) Seebeck coefficient as a function of electrical conductivity for SBTO+x wt% LSMO/LSCO composites.

A colossal value of α (≈ 319 mV/K) is observed at 300 K for the SBTO+30 wt% LSMO sample. Such colossal value of α at 300 K may be attributed to the quantum confinement of carriers as well as due to the presence of heavy carriers in the Aurivillius phase system [33]. Such confinement of carriers exhibits exotic transport properties due to variation of density of states (DOS) near the bottom of the conduction band and/or top of valence band with increasing confinement [45]. Cutler and Mott [46] showed that the electronic structure of metal and semiconductor could be described in terms of density of states (DOS), $\eta(E)$, as a function of energy E, and the dependence of α on $\sigma(E)$ is related as

$$\alpha = \frac{\pi^2 k_B^2 T}{3e} \left( \frac{d \log \sigma(E)}{dE} \right)_{E=E_f}$$

Where $E_f$ is the Fermi level, and $\sigma(E)$ is written as

$$\sigma(E) = \frac{e^2 \eta(E) \tau(E)}{m^*(E)}$$

From these equations, it is seen that α can be enhanced by increasing the DOS. A large value of $d\eta/dE$ occurs around the region of a sharp peak in the $\eta(E)$ vs. E plot. For one dimensional chain, the DOS peak occurs at the bottom and the top of the conduction and valence band, respectively. For a uniform 2D lattice, the DOS peak occurs in the middle of the band. Thus,

Hicks-Dresselhaus predicted that α of low dimensional materials could be enhanced without affecting other TE parameters. Such enhancement in the α for oxide superlattices were observed in different systems [47][48]. Based on the above interpretation, we propose an understanding of the enhanced α obtained in the present study as follows: The natural superlattice structure of $SrBi_4Ti_4O_{15}$ AP consists of perovskite slab of $(SrBi_2Ti_4O_{13})^{2-}$ with $(Bi_2O_2)^{2+}$ layer acting as an interslab region. The $(Bi_2O_2)^{2+}$ motifs form a planar net of oxygen atoms with $Bi^{3+}$ occupying in an alternating sequence above and below the perovskite slabs, forming a $BiO_4$ square pyramid. The nano-layered $(Bi_2O_2)^{2+}$ structure between the $(SrBi_2Ti_4O_{13})^{2-}$ layers in the unit cell may lead to a high α due to the confinement of charge carriers. Also, the charge carrier density for SBTO ($\approx 10^{13}$ $cm^{-3}$) and LSMO/LSCO ($\approx 10^{20}$ $cm^{-3}$) are different by a few orders of magnitude. This carrier concentration gradient across the interface of SBTO and LSMO/LSCO phases in the composite may also lead to enhanced α. However, a proper detailed theoretical investigation is required to confirm the presence of the colossal Seebeck coefficient in the AP system. A decrease in α with increasing temperature is observed and may be attributed to a decrease in the carrier concentration gradient with an increase in temperature. The α for SBTO+30 wt%LSMO at 800 K is about 8.14 mV/K, which is large compared to many TE materials investigated so far at this temperature [49]. However, an α of -28 mV/K is shown for another AP system ($Bi_2VO_{5.5}$) at 1050 K [33]. Such a colossal value of α is an exciting feature for the present system.

Further, the α decreases with an increase in LSMO wt% in the composite. The α at 300 K for the SBTO+50 wt%LSMO sample is ≈ 282 mV/K, and it also decreases with an increase in temperature. The same nature of α is observed for the SBTO+50 wt%LSCO sample. The Seebeck coefficient for SBTO+50 wt%LSCO is 268 mV/K at 300 K. The SPS samples show almost similar α (≈ 226 mV/K for SBTO+50 wt%LSMO-SPS and ≈ 224 mV/K for SBTO+50 wt%LSCO-SPS) at 300 K. The α for all the composite samples is found to decrease with increase in temperature showing the semiconducting nature of the composite.

The σ as a function of temperature (300 K-800 K) for all the composite samples is also shown in Fig. 3(a). The σ for the composite increases with the increase in LSMO and LSCO content, attributed to the increase of the conducting phase in the composite. It is worth noting that the electrical conductivity of the SBTO-LSCO composite is higher than the SBTO-LSMO composite due to the high electrical conductivity of LSCO compared to LSMO. Also, the pristine LSCO possesses a high Seebeck coefficient (~35 µV/K @300 K) compared to LSMO (~10 µV/K @300 K). The σ also increases with the increase in temperature, indicating a thermally activated behavior of the composite. A two-fold increase in the σ for SBTO+50 wt%LSMO-SPS and SBTO+50 wt%LSCO-SPS samples is observed compared to their non-SPS samples due to decrease in the grain boundaries in the SPS samples, as confirmed from FESEM images, which may scatter fewer charge carriers in the system.

Figure 3(b) depicts the Seebeck coefficient as a function of LSMO/LSCO weight fraction (x) in the composite. The Seebeck coefficient decreases with the increase in the weight fraction of the LSMO/LSCO phase in the composite. Also, Fig. 3(c) shows the change in the Seebeck coefficient as a function of electrical conductivity in the composite. The increase in electrical conductivity with the addition of the conducting phase decreases the Seebeck coefficient in the composite.

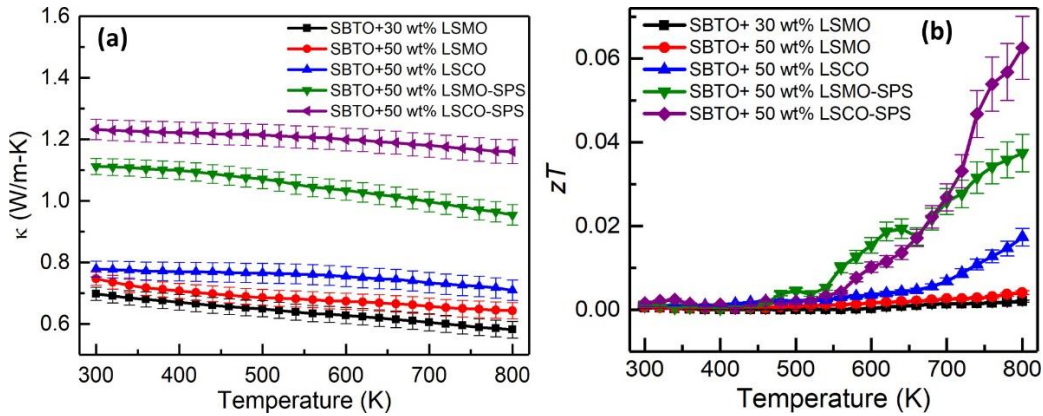

FIG. 4: (a) Thermal conductivity (κ) as a function of temperature and (b) Figure of merit zT as a function of temperature for SBTO+x wt% LSMO/LSCO composites. The solid line is a guide to the eye.

Next, the temperature-dependent thermal conductivity behavior of the composite is shown in Fig. 4(a). The κ for pure SBTO sample is ≈ 0.4 W/m-K at 300 K. This may be due to presence of $(Bi_2O_2)^{2+}$ interslab layer between $(SrBi_2Ti_4O_{13})^{2-}$ layer in the AP system which acts as a scattering center for phonon in the Aurivillius phase system itself [36]. Further, the κ for SBTO+50 wt%LSMO sample (≈ 0.75 W/m-K) is observed at 300 K. The mismatch in acoustic impedance between the SBTO-LSMO interface is expected due to their different sound velocity (2610 m/s for SBTO and 3170 m/s for LSMO), which leads to the acoustical mismatch in the composite [50]. However, with the increase in the LSMO/LSCO phase fraction, an increase in the κ is observed. As the κ in the composite is dominated by lattice thermal conductivity, a decrease in the κ at high temperatures may be attributed to the decrease in the mean free path at higher temperatures, which enhances the phonon scattering. In the SPS sample, almost twice an increase in the κ is observed and attributed to the reduction of grain boundaries, which reduces the phonon scattering [51]. The figure of merit ($zT = \alpha^2 \sigma T/\kappa$) for the composite samples is shown in Fig. 4(b). Despite having the colossal value of α and low κ for the SBTO+30 wt%LSMO sample at 300 K, the $zT$ is quite low and is due to the poor σ of the sample. At 800 K, the $zT$ value is found to increase with the increase in LSMO and LSCO addition in the composite. We observed a two-fold increase in the value of σ in SPS samples with a nominal decrease in the value of α, and this improves the $zT$ for SPS samples. The maximum $zT$ of 0.062 at 800 K is observed for the SBTO + 50 wt% LSCO-SPS sample.

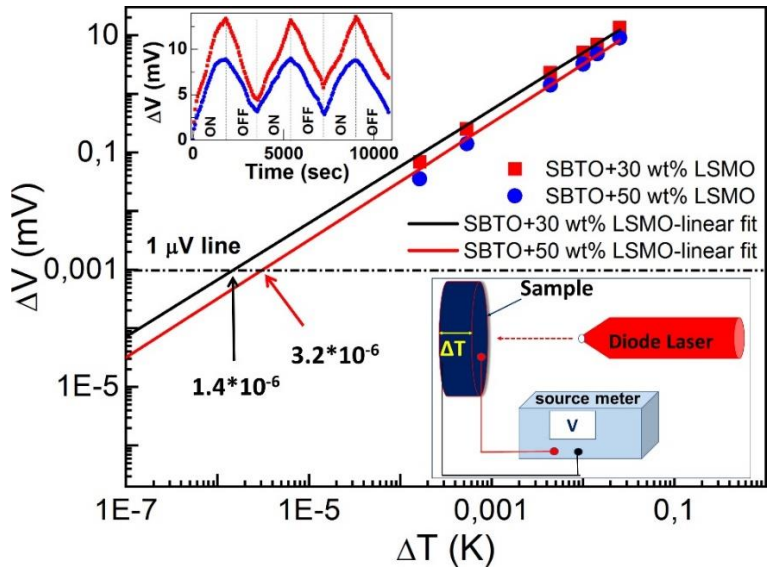

FIG. 5: Laser-induced change in the voltage (ΔV) of the composite samples as a function of change in temperature (ΔT) is shown. Inset shows the ΔV as a function of laser on-off time for three cycles for SBTO+30 wt% LSMO and SBTO+50 wt% LSMO composites. A schematic of the measurement is shown in the lower inset.

The colossal α, along with low κ, shows promise for application in thermopile sensing. In the present study, an external cavity diode laser (ThorLabs, λ=780 nm) of ≈7 mW power and 1 mm diameter spot size is used to observe the response of these composite materials to the laser-radiation at 300 K (schematic is shown in the inset of Fig. 5). A laser beam is allowed to fall on the sample (a cylindrical shape with 10 mm diameter and 3 mm thickness) via a pin-hole arrangement. This develops a ΔT, which results in a ΔV across the sample, as shown in Fig. 5. The change in the output voltage (ΔV) as a function of laser on-off time is shown in the upper inset of Fig. 5. For SBTO+30 wt% LSMO, the value of output voltage was found to be ~13.4 mV after ~40 minutes of laser irradiation, which corresponds to a ΔT of 0.042 K. Also, when the laser is switched off, the output voltage starts to decrease with time. The ΔV does not decrease to zero, as there is no heat sink used in this study, and this shows that there is some ΔT present across the sample. The on-off cycle was repeated three times, and almost the same response of the ΔV is observed. The ΔV obtained in the experiment was used to estimate the corresponding ΔT using the Seebeck coefficient value for each sample. The ΔV, as a function of ΔT, is plotted, as shown in Fig. 5. The behavior of ΔV vs. ΔT is found to be linear in log scale. We have extrapolated the ΔV vs. ΔT curve, and a 1 µV line was drawn. From the extrapolation curve, it has been found that a ΔV ~1 µV may be developed for a ΔT of 1.3-3.2 µK for the present SBTO-LSMO composite. This study suggests that these systems may be sensitive to tiny ΔT with straightforward synthesis techniques.

The colossal Seebeck coefficient obtained in the AP-perovskite oxide paves new directions for high-temperature thermoelectric materials. The LSMO and LSCO perovskite phases have been used as a conductive second phase in the AP composite materials. However, the ball-milled LSMO and LSCO with reduced particle size in the AP composite may lead to analogy to the brick-mortar model, which may be useful for further improvement in the $\alpha^2\sigma$ in the AP composites. Also, the size effect plays a vital role in the phonon-drag contribution to the Seebeck coefficient, as demonstrated by Yalamarthy et al. in the AlGaN/GaN 2DEG system at 300 K [52]. The nanocomposites approach enables the energy barrier scattering and improves the Seebeck coefficient, as demonstrated in several composites based on PbTe [18][19]. The results obtained in the present study in oxide composites using a simple synthesis strategy may be promising for future material innovations and potential use in high-temperature thermoelectric in the field of energy harvesting and room-temperature sensing applications.

## CONCLUSION

In summary, natural superlattice Aurivillius phase-perovskite composites are synthesized using a standard solid-state route. X-ray diffraction followed by two-phase Rietveld refinement confirms the pure phases present in the composites. Colossal α is observed in the SBTO-LSMO/LSCO composite sample and is attributed to the confinement of charge carriers in the unit cell. A two-fold increase in the σ is observed with the SPS samples at 800 K. The increase in σ results in a *zT* of 0.062 at 800 K for the SBTO+50 wt%LSCO-SPS sample. The colossal value of α enables this system to be used for thermopile sensors. The results obtained in this study paves further possibility to improve the *zT* in the AP system using substitutions at Bi and Ti sites. Also, the sensitivity of the thermopile sensor can be improved with further optimization in the geometry of the device.

## ACKNOWLEDGMENTS

The authors thank Prof. Raghwan K. Easwaran from IIT Patna, India for support with the diode laser facility and Prof. Jayant Kumar from University of Massachusetts Lowell, USA for fruitful discussion.